\documentclass[aps,preprint,showpacs]{revtex4}
\usepackage{graphicx}
\newcommand{\bce}{\begin{center}}
\newcommand{\ece}{\end{center}}
\newcommand{\be}{\begin{equation}}
\newcommand{\ee}{\end{equation}}
\newcommand{\bea}{\begin{eqnarray}}
\newcommand{\eea}{\end{eqnarray}}
\newcommand{\bit}{\begin{itemize}}
\newcommand{\eit}{\end{itemize}}
\newcommand{\E}{\>=\>}
\newcommand{\EA}{&=&}
\newcommand{\To}{\> \longrightarrow \> }
\newcommand{\non}{\nonumber \\}

\begin{document}

\vspace*{1cm}

\preprint{PSI-PR-03-04}

\vspace*{2cm}

\title{Structure Function of a Damped Harmonic Oscillator}

\vspace*{1cm} 
\author{R. Rosenfelder}
\email{roland.rosenfelder@psi.ch}

\affiliation{Particle Theory Group, Paul Scherrer Institute, CH-5232 Villigen PSI, 
Switzerland}

\date{\today}

\vspace*{1cm} 
\begin{abstract}
Following the Caldeira-Leggett approach to describe dissipative quantum systems the 
structure function for a harmonic oscillator with Ohmic dissipation is evaluated
by an analytic continuation from euclidean to real time. The analytic 
properties of the Fourier transform of the structure function with respect to the 
energy transfer (the ``characteristic function'') are studied and utilized. In the
one-parameter model of Ohmic dissipation we show explicitly that the broadening 
of excited states increases with the state number without violating sum rules.
Analytic and numerical results suggest that this is a 
phenomenologically relevant, consistent model to include the coupling of a single 
(sub-)nuclear particle to unobserved and complex degrees of freedom.
\end{abstract}

\maketitle

\section{Introduction}
The structure functions measured in deep inelastic lepton scattering provide
important information about the nature and momentum distribution of the 
constituents of the target. 
This is particularly true in the relativistic
domain where the point-like building blocks of matter, the quarks and gluons,
have been discovered and studied by inclusive scattering of multi-GeV electrons,
muons or neutrinos. 
In a non-relativistic description the (longitudinal) structure function for
one scalar particle is given by
\be
S(q, \nu) \E \sum_{\> n}\hspace{-0.5cm} \int \hspace{0.3cm} 
\delta(\nu - (E_n - E_0)) \, 
\left | < n | \exp (i q \cdot \hat x) | 0 > \right |^2 
\ee
where the summation is over all discrete and continuum states
which are excited by the probe. Here $q, \nu $ 
denote momentum and energy tranfer and $E_n$ the excitation energies of the 
target. Very often the confinement of quarks is described
by using rising potentials which lead to a purely discrete spectrum. 
The simplest version is given by a harmonic oscillator
potential which has a structure function \cite{HO}
\be
S^{\> \rm h.o.}(q, \nu) \E \sum_{n=0}^{\infty} \, 
\frac{\delta ( \nu - n \omega_0 )}{n !} 
\left ( \frac{q^2 b_0^2}{2} \right )^n \! \exp \left ( -  
\frac{q^2 b_0^2}{2} \right ) .
\label{S ho}
\ee
The oscillator frequency and length are denoted by 
$\omega_0$ and $ b_0 = 1/\sqrt{m \omega_0} $ where $m$ is the mass of the 
struck particle.
For simplicity, all many-body and recoil effects have been neglected
and the elastic line ($n = 0$) is included in the sum over excited states.
However, the observed structure functions are smooth due to
hadronization and/or final state interactions.
A number of recent theoretical studies 
have accounted for that by simply smearing out the $\delta$-functions
in eq.(\ref{S ho}) by a Breit-Wigner distribution with a constant width 
or averaged over nearby $\nu$-bins
\cite{smearing}. It is obvious, that this is not only {\it ad hoc} but also
may violate general properties of the structure function, 
e.g. the fact that it has to vanish below the first (positive) excitation 
energy.

It is the purpose of this paper to demonstrate that a consistent 
quantum mechanical framework exists which allows to treat couplings to 
unobserved degrees of freedom in a simple manner. Several methods 
have been used in the past to achieve that, for example 
for Coulomb excitation of particle-unstable states \cite{WeWi}.
For quasielastic scattering
of electrons from nuclei Horikawa {\it et al.} \cite{HLM} first have included 
multi-nucleon channels by employing an optical potential without violating the 
non-energy-weighted sum rule (NEWSR)
\be
\int_0^{\infty} d\nu \> S \left (q, \nu \right ) \E 1 \> .
\label{NEWSR}
\ee
However, there exists a simpler treatment based on 
the description of dissipative quantum systems within the path integral
formalism \cite{Weiss}. This originates in the celebrated work
of Feynman and Vernon \cite{FeVe} and Caldeira and Leggett \cite{CaLe} who have 
modelled the coupling of the system to an environment of $N (\to \infty)$
harmonic oscillators  
\be
H \E \frac{p^2}{2 m} + V(x) + \sum_{n=1}^{N} \, \left ( \, 
\frac{p_n^2}{2 m_n} + \frac{1}{2} m_n \omega_n^2 x_n^2 \, \right ) 
- x  \cdot \sum_{n=1}^N c_n x_n + x^2 \,   \sum_{n=1}^N  \, 
\frac{c_n^2}{2 m_n \omega_n^2}
\label{CL model}
\ee
with a bilinear coupling. The limit $N \to \infty $ of the number of  
environmental oscillators is essential in preventing the bounded motion 
of all particles to come back to the initial state after some time : the so-called
Poincar\'e recurrence time then tends to infinity \cite{HMW} and 
irreversibility becomes possible - still in a unitary quantum-mechanical framework. 
The infinite number of degrees of freedom also 
allows for strong damping even if each environmental oscillator couples 
only weakly to the system. This mechanism leads to a 
broadening (and a shift) of the $\delta$-functions 
in the structure function of the confined system without violating the sum rules.

In this approach the path integral description of the system offers particular 
advantages since the bath oscillators
can be integrated out exactly. This gives rise to a retarded two-time action for the
single particle which does not have a Hamiltonian counterpart anymore similar as 
in the time-honoured polaron problem \cite{pol}.
In particular, no Schr\"odinger equation for the single particle motion 
is available. However, if this particle moves in a harmonic potential
\be
V(x) \E \frac{1}{2} m \omega_0^2 \, x^2
\ee
then the remaining path integral can also be done exactly.
In section 2 we will employ this formalism and the explicit
results for the damped harmonic oscillator to obtain the Fourier transform 
of the structure function with respect to the energy transfer. As thermal physics 
lives in euclidean times one needs an analytic continuation to real times which
is performed in section 3. Numerical results are presented in section 4
while the conclusions are given in the final section. Some technical details
are collected in two appendices.

\section{Characteristic function and correlation functions of dissipative systems}

We will calculate the structure function from its Fourier transform
\be
S(q,\nu) \> =: \> 
 \frac{1}{2 \pi} \, \int_{-\infty}^{+\infty} dt \> 
e^{i \nu t} \, \Phi_q(t) \> ,
\label{S from Phi}
\ee
the ``characteristic function'' \cite{foot1}.
For the pure harmonic oscillator one has
\be
\Phi_q^{\rm h. o.}(t)  \E 
\exp \left [ - \frac{1}{2} q^2 b_0^2 \left ( 1 - e^{-i \omega_0 t}
\right ) \right ] 
\label{Phi ho}
\ee
which after expansion and integration leads to the result given in eq.
(\ref{S ho}).
For the damped harmonic oscillator the characteristic function can be related to
the particular correlation function in {\it euclidean} time $\tau$
\be
T(q,\tau) \E \left < 0 \left | {\cal T} \left ( e^{-i q \cdot \hat x(\tau)} \, 
e^{i q \cdot \hat x(0) } \right ) \right | 0 \right >  
\ee
by an analytic continuation.
From the spectral representations
\be
\left. \begin{array}{c} \Phi_q(t) \\
              T(q,\tau) \end{array} \right \} \E \sum_n \, | < n | \exp (i q 
\cdot \hat x) |0 > |^2 \, \cdot \, \left \{  \begin{array}{c}  
\exp \left [\,  - i ( E_n - E_0 ) t \, \right ] \\
           \exp \left [\, -  ( E_n - E_0 ) |\tau| \, \right ] \end{array} \right.
\ee
one sees that both expressions coincide for positive euclidean time and 
the correct analytic continuation is therefore obtained by considering 
$T(q,\tau > 0)$ and replacing $ \tau \to i t $.

In the path integral approach we have
\be
T(q,\tau) \E \lim_{\beta \to \infty} \frac{ \int {\cal D}x \, 
\exp [ -i q \cdot (x(\tau)-x(0)) ] \, \exp \{ - {\cal A} [x] \} }{
\int {\cal D}x \, \exp \{ - {\cal A} [x] \} } \> .
\label{Green func as path int}
\ee
Here $ {\cal A} [x] $ is the effective action of the particle after the 
oscillators of the environment have been integrated out and
the limit $\beta \to \infty $ of the final euclidean time 
projects out the ground state of the system \cite{ChLi}.
In this limit the boundary conditions for the path integrals 
in eq.(\ref{Green func as path int}) do not matter; 
therefore we may set $ x(-\beta/2) = x(\beta/2) = x$ and integrate over $x$, 
i.e. perform the thermodynamical trace. This allows us to directly take over 
results from dissipative quantum systems where similar correlation 
functions (e.g. for the position operator) have been evaluated at 
finite temperature, i.e. finite $\beta \> $ \cite{Weiss}. Since this 
is quite standard now we can be brief 
and immediately use results from the nice review by Ingold \cite{Ing}, in 
particular from chapter 4.3 with the 
driving force $F(\sigma) = i q \, [ \delta(\sigma-\tau) - \delta(\sigma) ] $. 
In the limit $ \beta \to \infty $ the sum over Matsubara frequencies 
turns into an
integral so that the final result for the euclidean correlation function reads
\be
T(q,\tau) \E \exp \left [ \, - \frac{q^2}{2 m} \, \frac{1}{\pi} 
\int_{-\infty}^{+\infty} dE \> \frac{1-\cos(E \tau)}{E^2  
+ |E| \, \gamma \left ( |E| \right ) + \omega_0^2} \, \right ] \> .
\label{T eucl}
\ee
Here 
\be
\gamma(E) \E \frac{2}{\pi m} \, \int_0^{\infty} d\omega \> \frac{J(\omega)}{\omega}
\, \frac{E}{E^2 + \omega^2}
\ee
is the damping kernel which is produced by the coupling 
of the system to the heat bath. One does not have to specify all masses, 
frequencies and coupling constants in Eq.(\ref{CL model}) but only the
spectral density $J(\omega) $ of the environment oscillators.
The simplest assumption is {\it Ohmic dissipation} 
\be
J_{\> \rm Ohm} \E m \gamma \omega \> \> \Longrightarrow \> 
\gamma_{\> \rm Ohm} (E) \E \gamma \> .
\ee
Although some observables (e.g. the ground state energy) need high-frequency
cut-offs \cite{foot2} this form can be used with impunity 
for the structure function where only energy differences matter. For simplicity
it will also be employed in the following. We then obtain
\be
T(q,\tau) \E \exp \left \{  \> - 2 q^2 \left [ \, 
x_V(0) - x_V(\tau) \, \right ] \right \} 
\ee
with 
\be
x_V(\tau) \E \frac{1}{2 m \pi} \int_0^{\infty} dE \, 
\frac{\cos(E \tau)}{E^2 + \gamma E + \omega_0^2} \> \> , \hspace{0.5cm} 
\left ( \, \tau \> {\rm real} \, \right ) \> .
\label{xv}
\ee
The analytic continuation of the above function which coincides with 
$x_V(\tau)$ for positive euclidean time
\be
\xi_V(\tau) \E x_V(\tau) \> \> , \> \> \left ( \, \tau \ge 0 \, \right )
\label{xiv 1}
\ee
may be called the {\it Vineyard function} since many years ago 
Vineyard \cite{Vine} derived 
an identical form of the structure function for inclusive scattering of 
slow neutrons from quantum liquids. Its nice feature is the clean separation
between the squared momentum transfer and the variable $t$ which is 
conjugate to the energy transfer.
Such a form was also 
obtained in a (zeroth order) variational calculation of relativistic
deep inelastic scattering from a scalar particle 
where the broadening of the 
elastic line was due to multiple meson production \cite{WC6}. 
We thus have found a simple description 
of inclusive scattering with one additional parameter $\gamma$ which accounts for 
the coupling of the particle to additional degrees of freedom  \cite{foot3}.
These could be the continuum
and/or many-nucleon emission in the case of quasielastic scattering 
from nuclei or the production of colorless hadrons in the case of scattering from
quarks.

Note that $T(q,\tau = 0) = 1 $ so that the sum rule (\ref{NEWSR})
is fulfilled. We will also see that the resulting structure function has the 
correct support, i.e. no unphysical excitations occur. This is because the 
Caldeira-Leggett model is based on a consistent many-body Hamiltonian and the
environmental degrees of freedom have been integrated out without approximation.
In contrast, other descriptions of the damped harmonic oscillator 
\cite{damped ho}, friction \cite{friction} or time asymmetry \cite{timeas}
in general require a modification of usual quantum mechanics.

\section{Analytic continuation}

The main task left over is to perform the analytic continuation 
to get the ``characteristic function'' $\Phi_q(t)$ from the correlation function
$T(q,\tau)$. This requires the analytic continuation of the Vineyard function
$\xi_V(\tau \to it)$.

It should be emphasized that $x_V(\tau)$ and $\xi_V(\tau)$ are {\it different}
functions which only coincide for $\tau \ge 0$. In particular, although 
from eq. (\ref{xv}) $x_V(\tau)$  is even in $\tau$ we will see that the 
Vineyard function
has a logarithmic cut on the negative real $\tau$-axis. In the following we
will always indicate the range of validity 
of the representation for the Vineyard function in parenthesis as was
done in eq. (\ref{xiv 1}). It is obvious that this form cannot be used for 
analytic continuation 
to Minkowski time (where scattering occurs) because the cosine-function 
would blow up. As shown in Appendix A we may, however, distort the integration 
contour as in ref. \cite{WC3} and obtain for $\tau > 0 $
\be
\xi_V(\tau) \E \frac{\gamma}{2 m \pi} \int_0^{\infty} dE \, 
\frac{E}{(E^2 - \omega_0^2)^2 + \gamma^2 E^2} \, e^{-E \tau} 
\> =: \> \int_0^{\infty} dE \, \rho(E)  \, e^{-E \tau} \> \> , 
\hspace{0.3cm} \left ( \, \mbox{Re} \> \tau \ge 0 \, \right ) \> .
\label{xiv 2}
\ee
This now allows an analytic continuation $ \tau \to i t $ to obtain the
characteristic function
 \be
\Phi_q^{\> \rm d.h.o.}(t) \E  \exp \left \{ \, - 2 q^2 \left [ \, \xi_V(0) - 
\xi_V(it) \, \right ] \, \right \} \> .
\label{Phi dho}
\ee
To see the effects of the damping we insert eq. (\ref{xiv 2}) 
into eqs. (\ref{Phi dho}),(\ref{S from Phi}) , expand the exponential and 
perform the $t$-integration. This gives
\bea
S^{\rm \> d.h.o.} (q,\nu) \EA   e^{- q^2 b^2/2} \, \delta(\nu) 
 + \sum_{n=1}^{\infty} \, \frac{(2 q^2)^n}{n!} \, e^{- q^2 b^2/2}  \non
&& \hspace{2cm} \cdot  \int_0^{\infty} dE_1 \ldots dE_n \, 
\delta \left ( \nu - \sum_{k=1}^n 
E_k \right ) \, \rho(E_1) \ldots \rho(E_n)  \> .
\label{S dho 1}
\eea
The first term is the elastic line with the square of the typical 
gaussian form factor for the harmonic oscillator. However, the oscillator 
length is renormalized by the interaction with the environment:
\be
b_0^2 \To b^2 \E 4 \, \xi_V(0) \E b_0^2 \> \, \frac{\omega_0}{\Omega} \,
\frac{2}{\pi} \arctan \left ( \frac{2 \Omega}{\gamma} \right ) \> \le\> b_0^2
\ee
with 
\be
\Omega = \sqrt{\omega_0^2 - \frac{\gamma^2}{4}} \>. 
\label{def Omega}
\ee
(we are only considering the
underdamped case $ \gamma < 2 \omega_0  $ ).
Compared with eq. (\ref{S ho}) all excited states are now broadened; in 
particular, the ($n=1$)-term just maps the weight function 
\be
2 q^2 \, e^{-q^2 b^2/2} \, \Theta(\nu) \, \rho(\nu)  \E 
\frac{q^2 b_0^2}{2} \,  e^{-q^2 b^2/2} \, \frac{2 \omega_0}{\nu+\omega_0} \,
\frac{\Theta(\nu)}{2 \pi} \frac{\Gamma(\nu)}{(\nu-\omega_0)^2 + 
\Gamma^2(\nu)/4} \> .
\ee
Apart from the additional factor $ \> 2 \omega_0/(\nu + \omega_0) \> $ the line 
shape is just an one-sided Breit-Wigner distribution with the 
energy-dependent width 
\be
\Gamma(E) \E \frac{2 E \gamma}{E + \omega_0} \> .
\ee
Note that this distribution vanishes at threshold and is identically zero for
unphysical negative energy transfers. The same holds for all other terms in 
the expansion (\ref{S dho 1}) due to the $\delta$-function and the fact that all 
$E_k \ge 0$. Thus the structure function of the damped harmonic oscillator 
has the correct support. Unfortunately, it is not possible to 
evaluate the higher-order terms analytically. Only in the 
{\it narrow-width approximation} 
\be
\Gamma(E) \> \approx \> \Gamma(\omega_0) \E \gamma 
\ee
a simple result is found 
when additionally the prefactor is also evaluted at 
$\nu = \omega_0$  and the $E$-integration extended to $-\infty$:
\be
\xi_V(it)  \> \approx \> \frac{b_0^2}{4}
\exp \left (  - i \omega_0 t - \frac{\gamma}{2} |t| \right ) \> .
\label{narrow}
\ee
Note that an {\it exact} expression of this form together with 
a real correction $r(t)$ is derived in Appendix \ref{app:A}
One then sees from  eqs. (\ref{Phi dho}),(\ref{S from Phi}) that
the $\delta$-function in the $n$-th excited state
contribution to the structure function 
(\ref{S ho}) of the harmonic oscillator turns into a Breit-Wigner function
with a width 
\be
\Gamma_n \> \approx \> n \gamma \> .
\ee
This result is well-known from the density of states of the
damped harmonic oscillator \cite{HaZw} and has also been discussed
for the width of  multi-phonon giant resonances \cite{giant}.
In the present context, however, it should be stressed that it is only approximate
and leads to a small, but non-vanishing structure function for $\nu < 0 $
\cite{foot4}.
This is due to the wrong analytic behaviour of the Vineyard function 
in the approximation (\ref{narrow}) where $ |t| = \sqrt{t^2} $  also 
produces a cut for Im $t < 0$.
In contrast, the exact expression has only a logarithmic cut in the upper 
half $t$-plane in accord with general properties of the 
characteristic function.
This can be best seen in the explicit expression of
the Vineyard function with Ohmic damping in terms of the standard exponential 
integral which is derived in Appendix A.
From eq. (\ref{A:xiv log}) it takes the form
\be
\xi_V(\tau) \E \mbox{regular function} \> - \> \frac{1}{2 m \pi \Omega} \, 
\sinh \left ( \Omega \tau \right )  \sin \left ( \frac{\gamma}{2} \tau \right )
\, \cdot \, \ln \left (\omega_0 \tau \right ) \> \> , \> \> | {\rm arg} \, \tau | 
< \pi 
\ee
for arbitrary complex $\tau$ away from the cut. Evaluating eq. (\ref{S from Phi}) 
for $\nu < 0$ by closing the integration contour 
in the lower half $t$-plane one therefore encounters no singularities and
the structure function vanishes identically.
The logarithmic cut of the Vineyard function also shows up in 
the low-$t$ expansion (see eqs. (\ref{A:xiv low tau}), (\ref{A:coeff}))
\bea 
\xi_V(it) \EA \xi_V(0) - \frac{i t}{4 m} + \frac{\gamma}{4 m \pi} \, t^2 
\ln ( i \omega_0 t ) \non
&& \hspace{2cm} - \frac{t^2}{4 m \pi \Omega} \, \left [ \, 
\left ( \Omega^2 - \frac{\gamma^2}{4} \right ) \arctan \left ( 
\frac{2 \Omega}{\gamma}
\right ) + \gamma \Omega \left ( \frac{3}{2} - \gamma_E \right ) \, \right ]
+ {\cal O}(t^3) 
\label{xiv small t}
\eea
which has important consequences: first, we see
that also the energy-weighted sum rule (EWSR)
\be 
\int_0^{\infty} d\nu \, \nu S(q, \nu) \E i \, \Phi_q'(0) \E \frac{q^2}{2m}
\label{EWSR}
\ee
is conserved \cite{foot5} whereas 
higher energy moments of the structure function diverge. This high-energy tail 
reflects, of course, the insufficient suppression of high frequencies in the 
simple model of Ohmic dissipation which may need modification for phenomenological 
applications in nuclear and quark physics. 
Second, it is even possible
to determine the tail for large energy transfer analytically: as
shown in Appendix B one obtains 
\be
S^{\rm \> d.h.o.}(q,\nu) \> \stackrel{\nu \to \infty}{\longrightarrow} \>
\frac{\gamma}{m \pi} \, \frac{q^2}{\nu^3} + \frac{3}{2} \,\frac{\gamma}{m^2 \pi} 
\, \frac{q^4}{\nu^4}  + {\cal O} \left ( \frac{\ln \nu}{\nu^5} \right ) \> .
\label{S asy}
\ee
Compared
with eq. (\ref{S dho 1}) one sees that the suppression of the low-lying states by
the square of the elastic form factor has disappeared and the asymptotic form
(\ref{S asy}) does not depend anymore on $\omega_0$ or the oscillator parameter
$b$ -- a property which roughly resembles  the conjectured 
``quark-hadron duality'' \cite{smearing}.

Another consequence of eq. (\ref{xiv small t}) is
that logaritmic corrections to $y$-scaling will persist even for large momentum 
transfer. This is because interaction times $ t \sim 1/q $ are probed in that 
limit \cite{RiRo} and therefore 
$ q^2 t^2 \ln \left (i \omega_0 t\right ) $ remains unbounded for $q \to \infty$.

\begin{figure}[ht]
\resizebox{0.62\textwidth}{!}{
\includegraphics{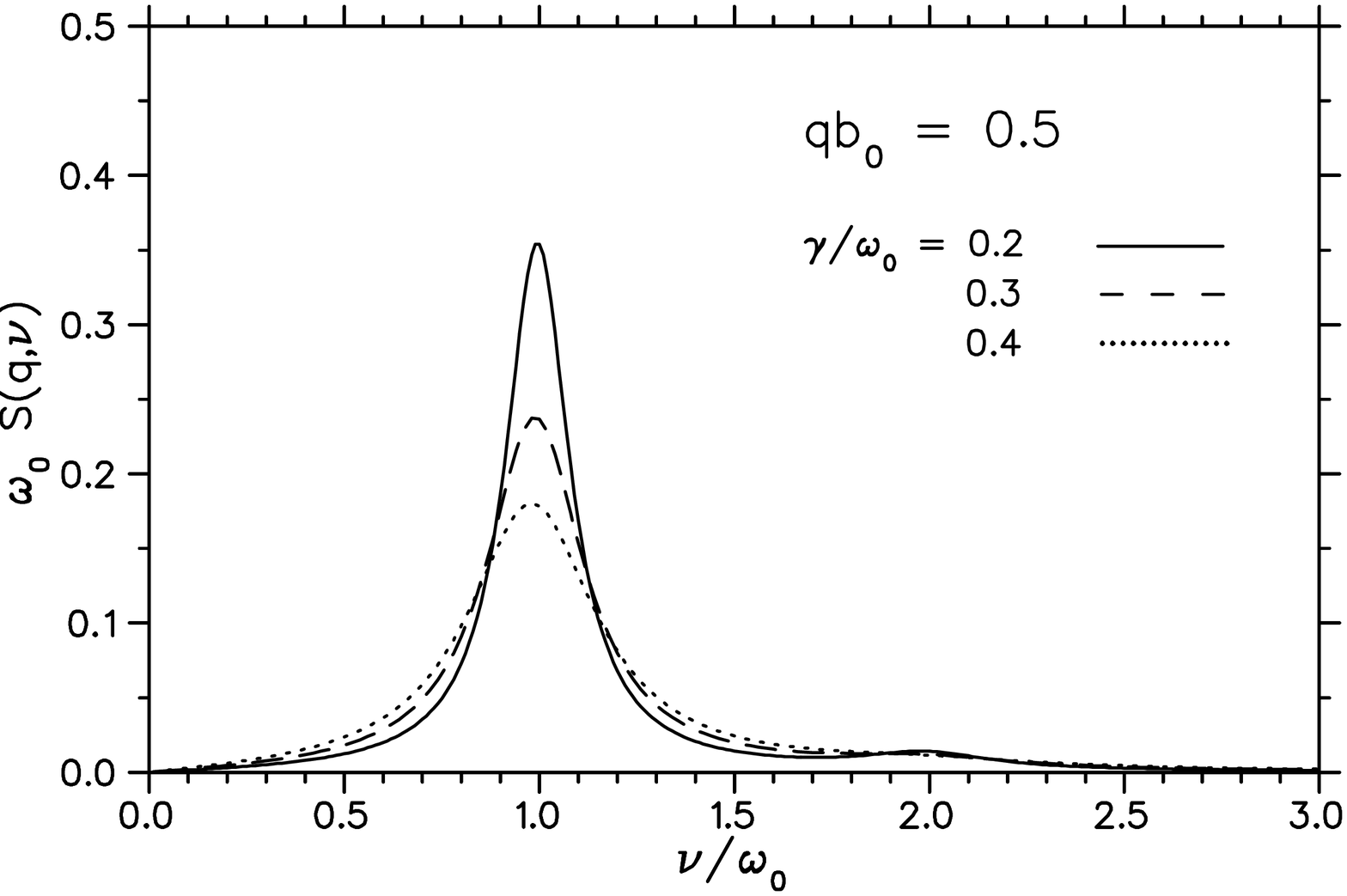}}
\\
\resizebox{0.62\textwidth}{!}{
\includegraphics{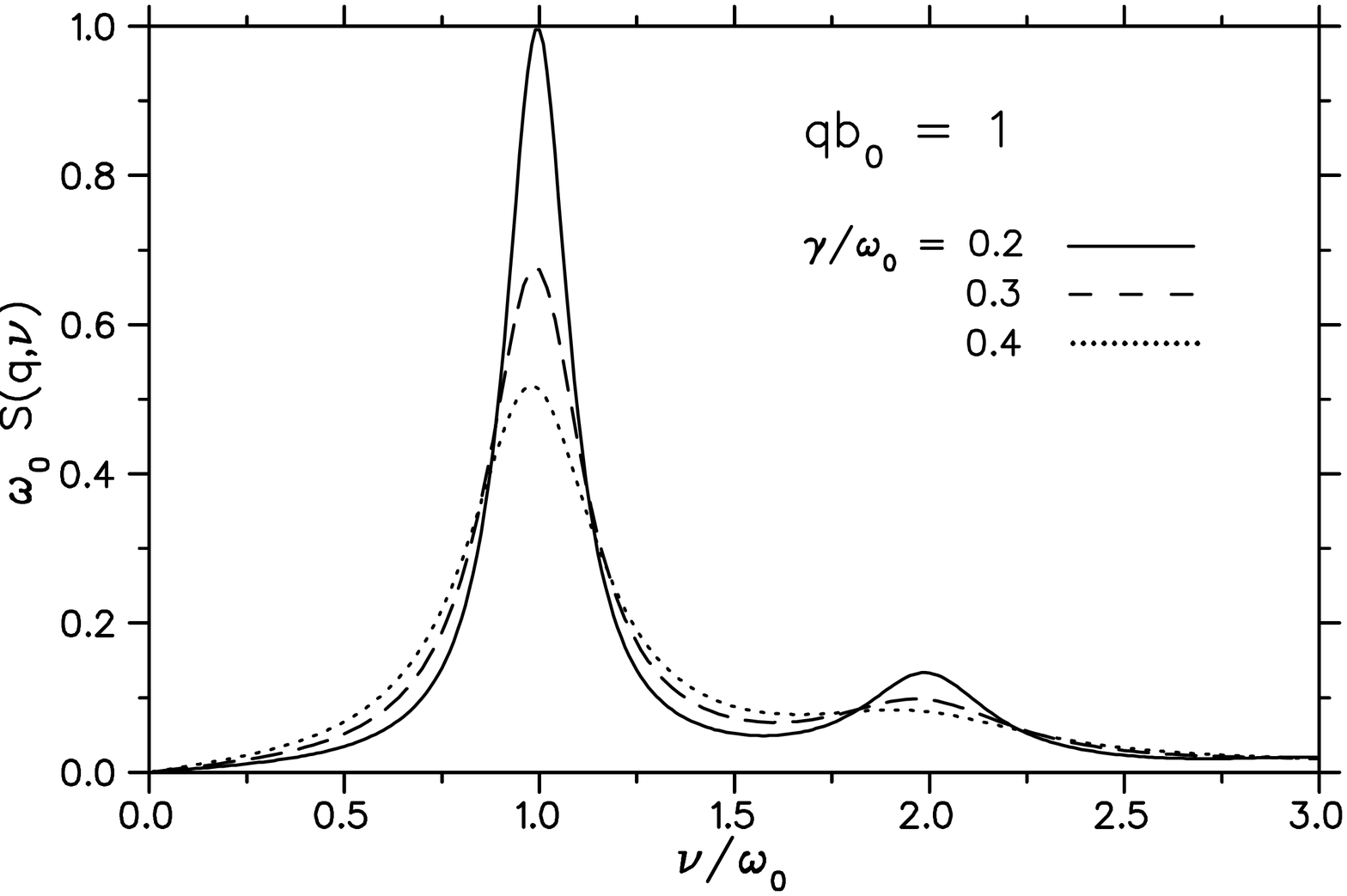}}
\caption{\label{fig: q05_10}The structure function of a damped harmonic 
oscillator as a function of the energy transfer for momentum transfers 
$q \,b_0 = 0.5 $ (top) and
$q \,b_0 = 1 $ (bottom). Note the different scales in both plots.
The solid curves are for a value of the Ohmic
damping parameter $\gamma/ \omega_0 = 0.2 $, the dashed one for  
$\gamma/ \omega_0 = 0.3 $ and the dotted one for  $\gamma/ \omega_0 = 0.4 $. 
The undamped oscillator length and frequency are denoted by 
$b_0$ and $\omega_0$, respectively.}
\end{figure}

\section{Numerical results}

Sticking to pure Ohmic dissipation we next try to evaluate the 
structure function quantitatively. Only for $q \to 0 $
(i.e. photoabsorption) the expansion (\ref{S dho 1}) into a sum over excited 
states 
is useful. For arbitrary momentum transfer the numerical problem is much harder
as one has to calculate the {\it inelastic} structure function 
\be
S^{\rm \> d.h.o.}_{\rm inelastic}(q,\nu) \E \frac{1}{2 \pi} \, 
\exp[ -2 q^2 \xi_V(0) ] \, \int_{-\infty}^{+\infty} dt \> e^{i \nu t} \, 
\left \{  \> \exp \left [ \, 2 q^2 \xi_V(it) \, \right ] - 1 \>  \right \}
\label{S inel 1}
\ee
after subtraction of the elastic line 
(a $\delta$-function) as a Fourier transform over an infinite interval.
One may alleviate the numerical problem slighthly by 
expressing the inelastic structure function as a sine
transform over the imaginary part of the characteristic function as demonstrated in 
ref. \cite{quasi}
\be
S^{\rm \> d.h.o.}_{\rm inelastic}(q,\nu) \E e^{- \frac{1}{2} q^2 b^2} \>  
\frac{2}{\pi} \int_0^{\infty} dt \> \sin(\nu t ) \> \left ( - {\rm Im} \right )  
\, \left \{ \, \exp \left [ \, 2 q^2 \, \xi_V(it) \, \right ] \, \right \} \> .
\label{S inel 2}
\ee
This holds since $\xi_V(it)$ vanishes as $1/t^2$
in a sector of the complex $t$-plane which includes the lower 
half-plane (see Appendix A). Thus one may write a Cauchy integral representation
for the (inelastic) characteristic function and express its real
part in terms of the imaginary part. 
We have used the adaptive integration routine D01ASF from the NAG library 
together with the explicit representations (\ref{Im + r(t)}), (\ref{r(t) 3}) 
of the Vineyard function to perform the numerical evaluation of eq. 
(\ref{S inel 2}).
 
\begin{figure}[th]
\resizebox{0.62\textwidth}{!}{
\includegraphics{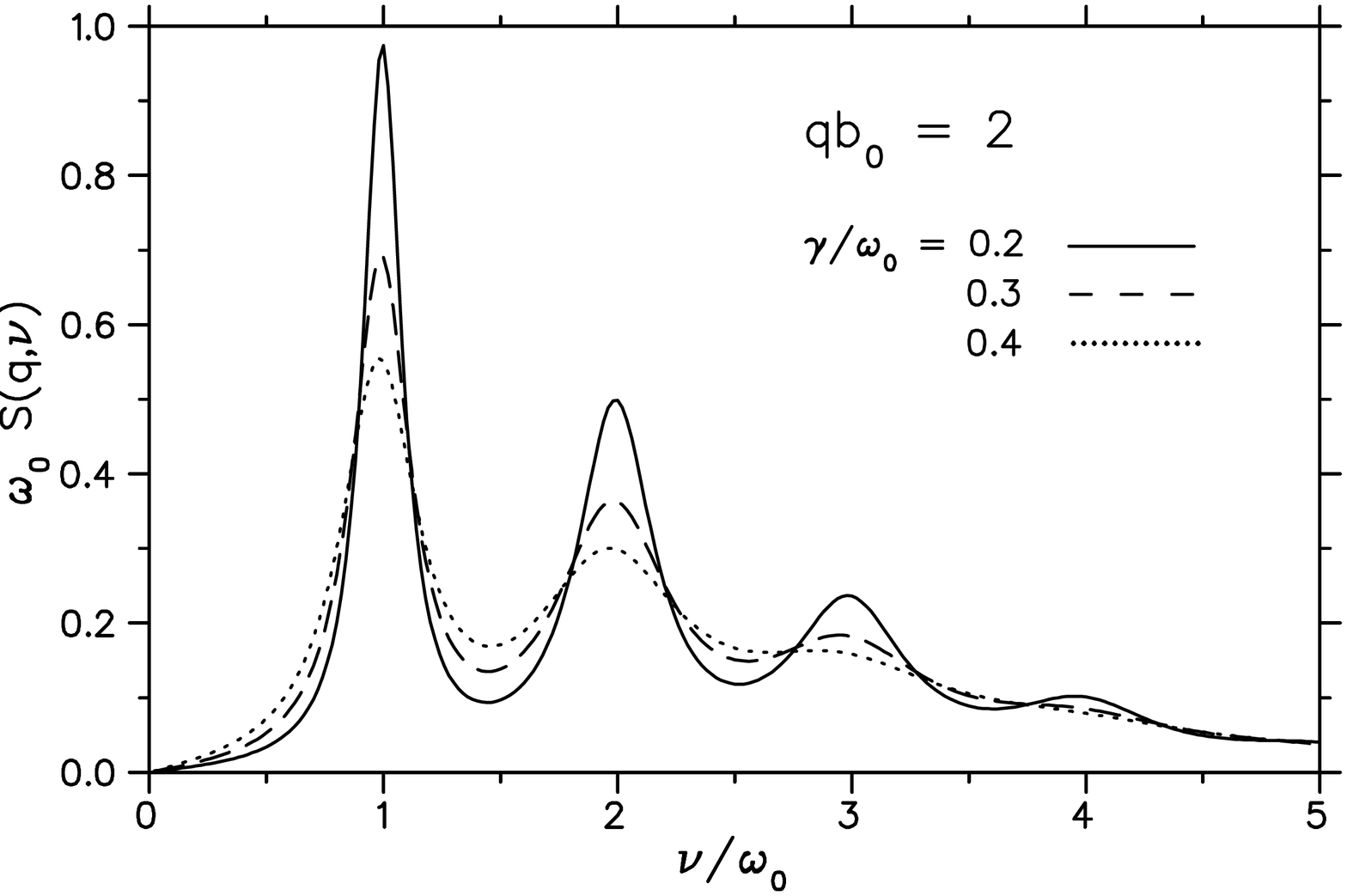}}
\\
\resizebox{0.62\textwidth}{!}{
\includegraphics{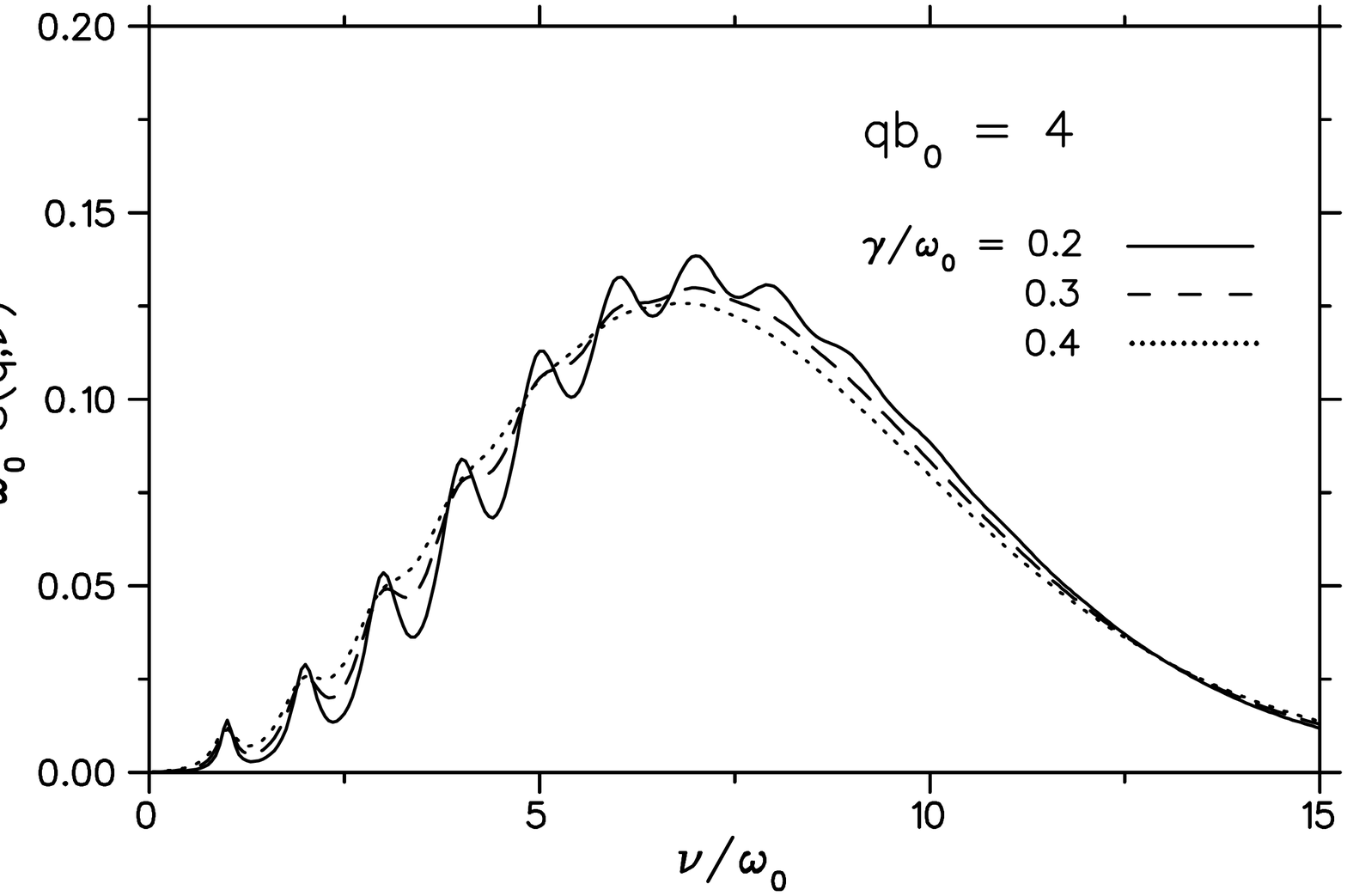}}
\caption{\label{fig: q20_40}Same as in Fig. \protect{\ref{fig: q05_10}} 
but for $q \, b_0 = 2$ (top) 
and $q \, b_0 = 4$ (bottom).}
\vspace{-0.1cm}
\end{figure}

Figs. \ref{fig: q05_10} and \ref{fig: q20_40}
show the results of the calculation for several momentum transfers and
damping parameters. It is seen that the excitation of individual levels gradually
moves into the broad structure of the quasi-elastic peak as the momentum transfer
increases. The value $\gamma/\omega_0 = 0.2 $ corresponds roughly to the one
used in ref. \cite{resp} where a 
parametrization of the response function was fitted to photoabsorption 
and electron scattering data in $^{12}$C. A  peak position of $22.7$ MeV and a 
FWHM of $ (2 \sqrt{\ln 2} \cdot 2.6 =
4.3) $ MeV was obtained for the giant resonance.
Although accounting for the coupling 
of this excitation to many-body states and the continuum 
this value of $\gamma/\omega_0$ gives too much 
structure from individual levels at higher $q$ compared with typical experimental
cross section. This may indicate that a modification of the assumed Ohmic damping 
is needed for a description of medium-energy 
inclusive scattering data from nuclei. 

\noindent
That the sum rules are well preserved
can be seen in Tables \ref{tab:NEWSR}, \ref{tab:EWSR} where the relative size 
of different contributions is
presented for $\gamma/\omega_0 = 0.2 $ (other values of the damping 
parameter yield similar results). The numerical part was obtained by 
integrating the structure functions 
shown in Figs. \ref{fig: q05_10}, \ref{fig: q20_40} with weight 
$\nu^n \, , \, n = 0, 1 $ from $ \nu = 0 $ to 
$ \nu = \nu_{max} $ in steps of $\Delta \nu$ by means of a Simpson rule.
The NEWSR, of course, also gets a contribution $ \exp ( - q^2 b^2/2) $
from the elastic line. Also listed are the asymptotic contributions
obtained from eq. (\ref{S asy}) by integrating from $\nu_{max}$ to $\infty$.
It is seen that the asymptotic contributions substantially improve the 
convergence to the correct values except for a few cases at high momentum where  
$\nu_{max}$ is not large enough.

Due to its weighting the EWSR is, of course, not as well 
fulfilled as the NEWSR, but the agreement is very satisfactory and sufficient to
demonstrate the consistency of {\it all} parts of the calculation: elastic line, 
inelastic excitations and asymptotic behaviour of the structure function.
\begin{table}[htb]
\begin{tabular}{|c|c|c|c|c|c|c|c|} \hline 
\quad $q \,b_0$  \quad & \quad $\nu_{max}/\omega_0$ \quad & \quad 
$\Delta \nu/\omega_0$ \quad & \quad \quad elast. \quad \quad & \quad \quad num. 
\quad  &
\quad asy. (LO) \quad & \quad asy. (NLO) \quad & \quad total \quad \quad  
\\ \hline 
 0.5     &        3. & 0.01   & 0.88904 & 0.10966 & 0.00088 & 0.00007 & 0.99965 \\
         &        4. &        &         & 0.11037 & 0.00050 & 0.00003 & 0.99994  \\
         &        5. &        &         & 0.11061 & 0.00031 & 0.00002 & 0.99998  \\
 1.0     &        3. &        & 0.62471 & 0.36250 & 0.00354 & 0.00118 & 0.99193 \\
         &        5. &        &         & 0.37334 & 0.00127 & 0.00025 & 0.99957 \\
         &        7. &        &         & 0.37448 & 0.00065 & 0.00009 & 0.99993 \\
 2.0     &        5. & 0.02   & 0.15230 & 0.80314 & 0.00509 & 0.00407 & 0.96460 \\
         &        8. &        &         & 0.84239 & 0.00200 & 0.00099 & 0.99768 \\
         &      10.~ &        &         & 0.84532 & 0.00127 & 0.00051 & 0.99940 \\
 4.0     &      15.~ & 0.05   & 0.00054 & 0.97227 & 0.00226 & 0.00241 & 0.97749 \\
         &      20.~ &        &         & 0.99455 & 0.00127 & 0.00102 & 0.99738 \\
         &      25.~ &        &         & 0.99744 & 0.00082 & 0.00052 & 0.99932 
\\ \hline
\end{tabular}
\vspace{0.2cm}
\caption{Test of the non-energy-weighted (NEWSR) sum rule (\protect{\ref{NEWSR}}) 
for different momemtum transfers $q$ and
upper limits $\nu_{max}$ of the numerical integration performed in steps of 
$\Delta \nu$. The parameter for Ohmic damping is taken as $\gamma/\omega_0 = 0.2$.
The columns labelled ``elast.''  and ``num.'' denote the contribution 
of the elastic line and the result from numerical integration up to 
$\nu = \nu_{max}$, respectively. The columns ``asy.'' list the leading-order 
(LO) and next-to-leading (NLO) asymptotic
contributions from $\nu_{max}$ to infinity. In places with no entry the previous
value applies.}
\vspace{0.5cm}
\label{tab:NEWSR}
\end{table}

Although the NAG routine D01ASF does an excellent job in evaluating the
oscillating integral over the characteristic function one may ask whether it is
possible to introduce additional damping by deforming the integration contour
in eq. (\ref{S inel 1}) in the upper-half $t$-plane. Indeed, running along 
both sides of the cut would eliminate the oscillating exponential factor 
altogether.
\begin{table}[htb]
\begin{tabular}{|c|c|c|c|c|c|c|} \hline 
\quad $q \,b_0$  \quad & \quad $\nu_{max}/\omega_0$ \quad & \quad 
$\Delta \nu/\omega_0$ \quad & \quad \quad num. \quad \quad & \quad asy. (LO) \quad & 
\quad asy. (NLO) \quad & \quad total \quad \quad \\ \hline
 0.5     &        3. & 0.01   & 0.9448 & 0.0424 & 0.0027 & 0.9899 \\
         &        4. &        & 0.9638 & 0.0318 & 0.0015 & 0.9971  \\
         &        5. &        & 0.9723 & 0.0255 & 0.0009 & 0.9987  \\
 1.0     &        3. &        & 0.8882 & 0.0425 & 0.0106 & 0.9413  \\
         &        5. &        & 0.9653 & 0.0255 & 0.0038 & 0.9946 \\
         &        7. &        & 0.9785 & 0.0182 & 0.0020 & 0.9987 \\
 2.0     &        5. & 0.02   & 0.8512 & 0.0255 & 0.0153 & 0.8920 \\
         &        8. &        & 0.9668 & 0.0159 & 0.0060 & 0.9887 \\
         &      10.~ &        & 0.9797 & 0.0127 & 0.0038 & 0.9962 \\
 4.0     &      15.~ & 0.05   & 0.9356 & 0.0085 & 0.0068 & 0.9509 \\
         &      20.~ &        & 0.9819 & 0.0064 & 0.0038 & 0.9921 \\
         &      25.~ &        & 0.9898 & 0.0051 & 0.0024 & 0.9973 \\ \hline
\end{tabular}
\vspace{0.2cm}
\caption{Same as in Table \protect{\ref{tab:NEWSR}} but for
the energy-weighted (EWSR) sum rule (\protect{\ref{EWSR}}) divided by $q^2/(2m)$. 
Note that no elastic contribution exists in this case.
The Ohmic parameter is again $\gamma/\omega_0 = 0.2$.}
\label{tab:EWSR}
\vspace{0.5cm}
\end{table}
However, this is not possible since as shown in Appendix A there are {\it Stokes}
lines in the upper-half $t$-plane which
separate the power-like decrease of the 
Vineyard function $\xi_V(it)$ from an exponential increase which would overwhelm
the exponential damping from the factor $ \exp (i \nu t) $. The optimal damping
which is achievable without contributions from the arcs at infinity is a rotation
by an angle
\be
\varphi_1 \E \arctan \left ( \frac{\gamma}{2 \Omega} \right )
\label{def phi1}
\ee
\begin{figure}[ht]
\vspace{0.5cm}
\resizebox{0.7\textwidth}{!}{
\includegraphics{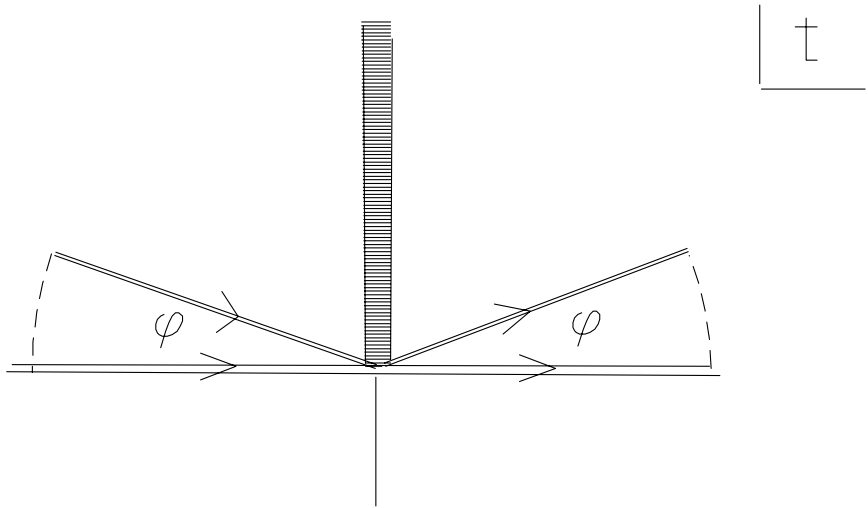}}
\caption{\label{fig:contour rot}Integration contours in the complex $t$-plane 
for the
numerical evaluation of the structure from the characteristic function: 
original path along the real axis (solid line) and path rotated 
by an angle $\varphi$ into the upper half-plane (double line). The arcs at 
infinity (dashed curves) do not give a contribution as long as 
$\varphi < \varphi_1$
where $\varphi_1$ is defined in eq. (\protect{\ref{def phi1}}). The 
logarithmic cut along the positive imaginary axis is shown as hatched strip.}
\vspace{0.3cm}
\end{figure}
which is depicted in Fig. \ref{fig:contour rot}. We have evaluated
\be
S^{\rm \> d.h.o.}_{\rm inelastic}(q,\nu) \E e^{- \frac{1}{2} q^2 b^2} \>  
\frac{1}{\pi} \, {\rm Re} \, \left ( \,  e^{i \varphi} \, \int_0^{\infty} dx \> 
\exp \left [ i \nu x  e^{i \varphi} \right ]  
\, \Bigl \{ \, \exp \left [ \, 2 q^2 \, \xi_V \left ( \tau = i x e^{i \varphi} 
\right ) \, \right ] \, - 1 \, \Bigr \} \, \right ) 
\label{S inel 3}
\ee
by standard Gaussian integration after mapping the infinite interval to a finite 
one. As expected the damping factor $\exp ( - \nu x \sin \varphi ) $ is most 
beneficial for large energy transfer whereas a greater
number of integration points is needed to avoid a negative numerical result
for the structure function at small $\nu$. Thus the contour rotation
method is complementary to an explicit summation over excited lines as given 
in eq. (\ref{S dho 1}).
 
\vspace{0.3cm}

\section{Summary}

We have shown that the Caldeira-Leggett model of the damped harmonic
oscillator also provides a consistent model for inclusive processes where it 
accounts for the coupling
of a single particle to more complicated states and additional degrees of freedom.
With only one additional (damping) parameter
in the simplest version with purely Ohmic dissipation 
a highly non-trivial structure function has been obtained 
which conserves the sum rules. Its characteristic function has been given in 
closed analytic form with a clear separation between momentum transfer and  
``time'', the variable conjugate to the energy transfer in the process. 
This allowed us to concentrate on the study of a function of a single variable 
-- the Vineyard function -- and its analytic properties which determine the 
dynamics. As an extension to the many-body case is straight-forward further 
applications in nuclear or hadronic physics seem to be possible.

\vspace{1cm} 
\noindent
\begin{acknowledgments}
This work originated from a course ``Path Integrals in 
Quantum Physics'' given in the winter semester 2002/3 at ETH Z\"urich. I am 
grateful to all students who forced me to think about problems which I hadn't 
considered before. Many thanks also to Gerhard Baur (J\"ulich) who pointed out 
relevant work on damping in nuclear reactions and multi-phonon giant resonances.
\end{acknowledgments}

\vspace*{2cm}

\noindent
\appendix
\section{The Vineyard function for the harmonic oscillator with Ohmic damping}
\setcounter{equation}{0}
\label{app:A}

Here we derive and collect a few properties of the Vineyard function defined by
eqs. (\ref{xiv 1}, \ref{xv}) in euclidean time $\tau \ge 0$
\be
\xi_V(\tau) \E \frac{1}{2 m \pi} \int_0^{\infty} dE \, 
\frac{\cos(E \tau)}{E^2 + \gamma E + \omega_0^2} \> .
\label{A:xiv 1}
\ee
One may 
decompose the cosine function into exponentials and deform the integration path 
such that it runs along the imaginary axes. There is no contribution from
poles of the integrand (which are all in the left-hand $E$-plane) and 
one therefore obtains (first for $\tau \ge 0 $)
\be
\xi_V(\tau) \E \frac{\gamma}{2 m \pi} \int_0^{\infty} dE \> 
\frac{E}{(E^2 - \omega_0^2)^2 + \gamma^2 E^2} \,  e^{- E \tau} 
\> \> , \> \> \left ( \, 
{\rm Re} \> \tau \ge 0 \, \right ) \> .
\label{A:xiv 2}
\ee
This is suitable for analytic continuation $\tau \to i t $ since the integral
also converges for $ {\rm Re} \> \tau \ge 0 $.
Using partial fractions one gets
\bea
\xi_V(\tau) \EA \frac{1}{8 m \pi i \Omega} \sum_{r,s = \pm 1} \, r s \, 
\int_0^{\infty} dE \>  \frac{1}{E + E_{r,s}} \,  e^{- E \tau} \non
\EA \frac{1}{8 m \pi i \Omega} \sum_{r,s = \pm 1} \, r s \, e^{E_{r,s} \tau} \, 
E_1 \left ( E_{r,s} \tau \right ) \> \> , \> \tau \ge 0 
\label{A:xiv 3}
\eea
with 
\be
 E_{r,s} \E r \Omega + i s \frac{\gamma}{2} \> \> , \> \> \> r, s = \pm 1 \> .
\label{def Ers}
\ee 
Here
\be
E_1(z) \E   - \gamma_E - \ln z - \sum_{n=1}^{\infty} \, \frac{(-z)^n}{n n!} 
\> =: \> - \gamma_E - \ln z + {\rm Ein}(z)
\ee
is the standard exponential integral and $ \gamma_E = 0.57721566 ..$
Euler's number.
A careful evaluation of the arguments of the logarithm  for $\tau \ge 0 $ 
then gives
\be
\xi_V(\tau) \E \frac{1}{8 m \pi \Omega} \, \sum_{r,s = \pm 1} \, 
e^{ E_{r,s} \tau } \, \left \{  \, \arctan \left ( 
\frac{2 \Omega}{\gamma}\right ) - r \frac{\pi}{2} 
+ i \, r s \, \Bigl [ \gamma_E + \ln (\omega_0 \tau) - {\rm Ein}
\left (  E_{r,s} \tau \right ) \, \Bigr ] \, \right \} \> ,
\label{A:xiv 4}
\ee
which defines the Vineyard function for {\it arbitrary} complex $\tau$ with
$ | {\rm arg}\, \tau |< \pi $. Since 
Ein($z$) is an entire function  (ref. \cite{Handbook}, chapter 5.1, footnote 3)
one sees that 
\be
\xi_V(\tau) \E \mbox{regular function} \> - \> \frac{1}{2 m \pi \Omega} \, 
\sinh \left ( \Omega \tau \right )  \sin \left ( \frac{\gamma}{2} \tau \right )
\, \cdot \, \ln \left (\omega_0 \tau \right ) \> .
\label{A:xiv log}
\ee
which allows to determine the discontinuity across the cut.

\noindent
For purely
imaginary arguments $\tau = i t \>,  t $ real, eq. (\ref{A:xiv 4}) can be 
written as
\bea
\xi_V(it) \EA \frac{1}{4 m \Omega} \, \exp \left ( - i \Omega t - 
\frac{\gamma}{2} |t| \right ) - r(t) 
\label{Im + r(t)}\\
r(t) \EA \frac{1}{4 m \pi \Omega} \, {\rm Im} \, \Bigl [ \> e^{ z_{\gamma}(t) } 
\, E_1 \left ( z_{\gamma}(t)
\right)   - ( \gamma \to -\gamma) \, \Bigr ]  \> \> , 
z_{\gamma}(t) \> \equiv \> 
\left ( i \Omega + \frac{\gamma}{2}\right ) |t| \> .
\label{r(t) 1}
\eea
The first term corresponds to the narrow-width approximation (\ref{narrow}) 
whereas the last term (which is real, even in $t$ and 
vanishing for $\gamma = 0$ ) corrects for its deficiencies. 

\begin{figure}[ht]
\resizebox{0.85\textwidth}{!}{
\includegraphics{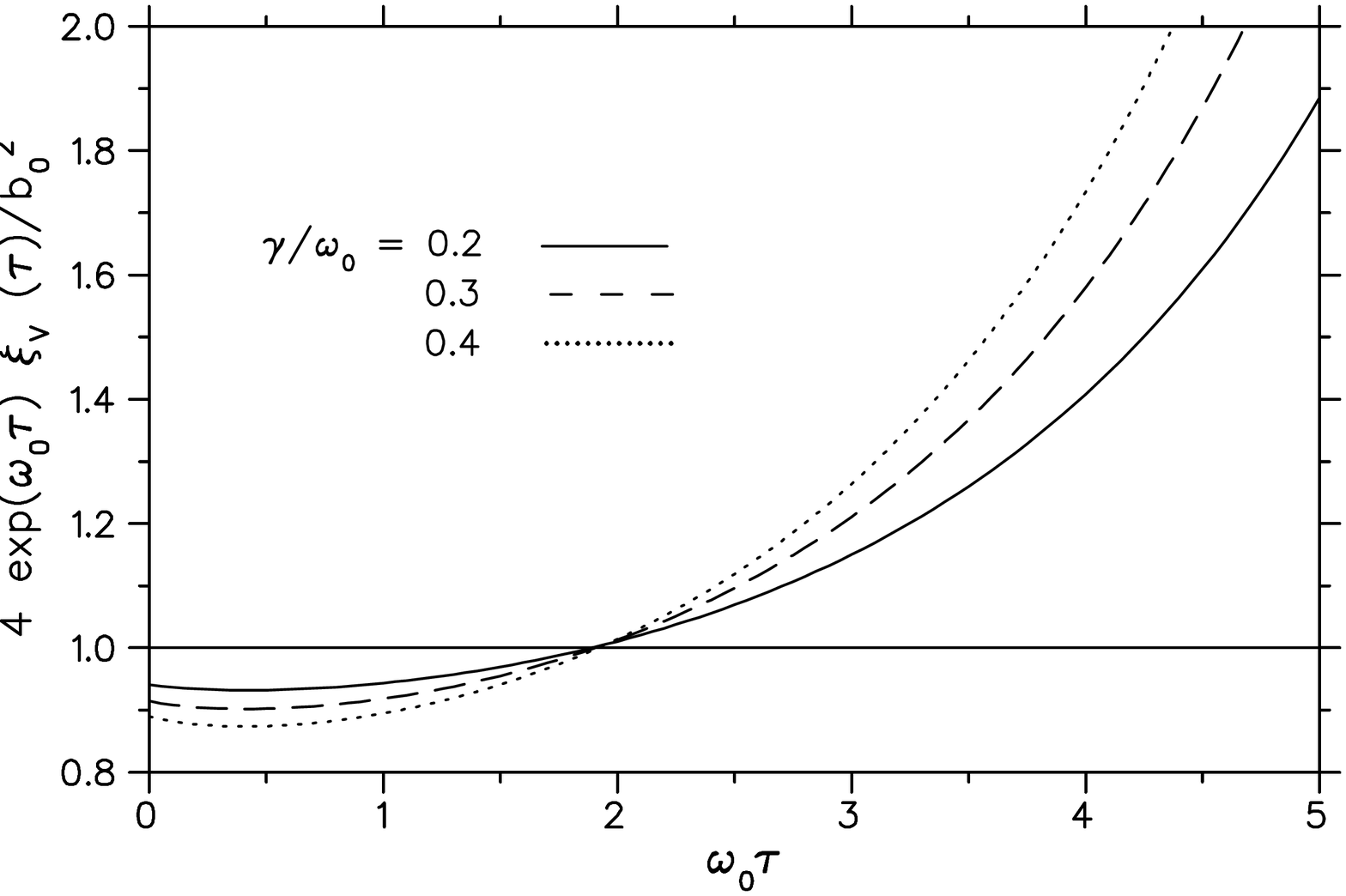}}
\caption{\label{fig:xi(tau)}The Vineyard function $\xi_V(\tau)$ normalized 
to the undamped case as function of the euclidean
time $\tau$. The solid curve is for a value of the Ohmic
damping parameter $\gamma/ \omega_0 = 0.2 $, the dashed one for  
$\gamma/ \omega_0 = 0.3 $ and the dotted one for  $\gamma/ \omega_0 = 0.4 $.}
\vspace{0.3cm} 
\end{figure}

\begin{figure}[ht]
\resizebox{0.85\textwidth}{!}{
\includegraphics{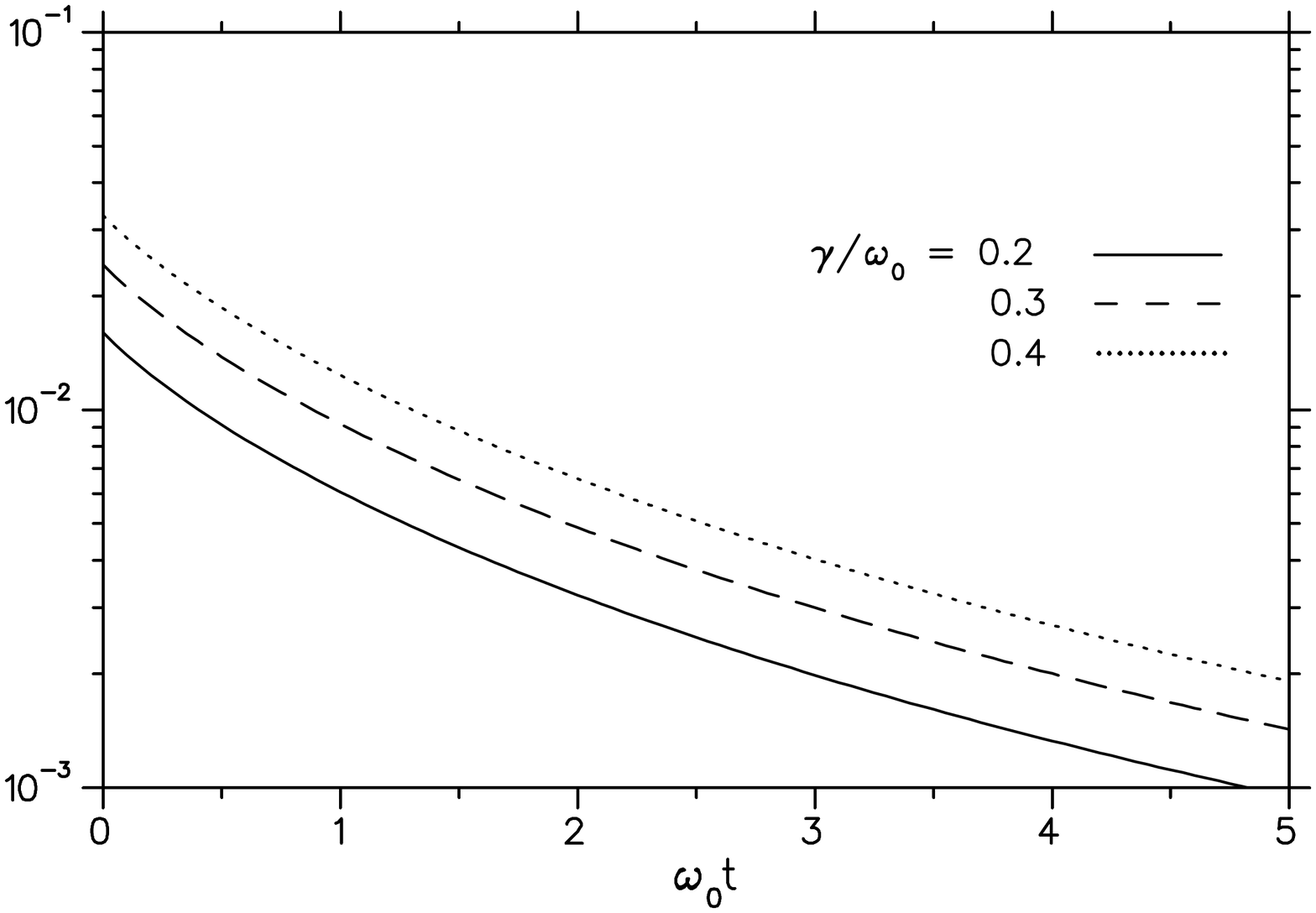}}
\caption{\label{fig:r(t)}The correction term $r(t)$ to the narrow-width 
approximation (\protect{\ref{Im + r(t)}}) for the Vineyard function $\xi_V(it)$ 
as function of the time $t$. Note the logaritmic scale for $r(t)$.}
\end{figure}

\noindent
Another representation 
of the remainder function $r(t)$ is obtained by using the identity
\be
\frac{E}{(E^2-\omega_0^2)^2 + \gamma^2 E^2} \E \frac{1}{4 \Omega} \, 
\left [ \, 
\frac{1}{(E-\Omega)^2 + \gamma^2/4} - \frac{1}{(E+\Omega)^2 + \gamma^2/4} \, 
\right ]
\label{antisym BW}
\ee
which shows that the weight function $\rho(E)$ can also be considered as an 
(anti-)symmetrized Breit-Wigner distribution around $ E = \Omega$ . 
Inserting this into eq. (\ref{A:xiv 2})  one obtains for $\tau = i t , t $ real
\bea
\xi_V(i t) \EA \frac{\gamma}{8 m \pi \Omega} \, \Biggl \{ \, 
\int_{-\infty}^{+\infty} 
dE \, \frac{\exp(- i E t)}{(E-\Omega)^2 + \gamma^2/4} -  \int_{-\infty}^0 
dE \, \frac{\exp(- i E t)}{(E-\Omega)^2 + \gamma^2/4} \non
&& \hspace{6cm} -  \int_0^{\infty} 
dE \, \frac{\exp(- i E t)}{(E+\Omega)^2 + \gamma^2/4} \, \Biggr \} \non
\EA  \frac{\gamma}{8 m \pi \Omega} \, \left \{ \, \frac{2 \pi}{\gamma} \, 
\exp \left ( - i \Omega t - 
\frac{\gamma}{2} |t| \right )  - 2 \int_0^{\infty} 
dE \> \frac{\cos (E t)}{(E+\Omega)^2 + \gamma^2/4} \, \right \} \> .
\eea  
Hence
\be
r(t) \E \frac{\gamma}{4 m \pi \Omega} \, \int_0^{\infty} 
dE \> \frac{\cos (E t)}{(E+\Omega)^2 + \gamma^2/4} \> \> , \> \> t \> 
\mbox{real} \>.
\label{r(t) 2}
\ee
Again one may distort the integration path such that it runs along the 
imaginary axis.
Decomposing the cosine function into exponentials and realizing that the 
integrand in eq. (\ref{r(t) 2}) has only poles at $E = - \Omega \pm i \gamma/2$, 
one then obtains a representation
\be
r(t) \E \frac{\gamma}{2 m \pi} \, \int_0^{\infty} 
dE \> \frac{E}{(E^2-\omega_0^2)^2 + 4 \Omega^2 E^2} \, e^{- E |t|}
\> \> , \> \> t \> \mbox{real} 
\label{r(t) 3}
\ee
which is very well suited for numerical evaluation. Indeed, we have checked the
routine which calculates the Vineyard function $\xi_V(it)$ based on
on the exponential integral representation (\ref{r(t) 1})
by a direct Gaussian integration of  eq. (\ref{r(t) 3}) and found 
a relative deviation 
\be
\left | \frac{\xi_V^{\rm expon. int.}(it) -  
\xi_V^{\rm num.}(it)}{\xi_V^{\rm num.}(it)}\right |  <  2 \cdot 10^{-6}
\ee 
for all real $t$ and $\gamma/\omega_0 \le 0.4$. In this comparison
the complex exponential integral $E_1(z)$ was calculated by the rational
approximations with $n = 10$ terms given in ref. \cite{Luke}: Table 64.4 was used
for $|z| < 9$ and Table 64.5 for $ |z| \ge 9$ and checked against values
listed in Table 5.6 of ref. \cite{Handbook}. Fig. \ref{fig:xi(tau)}
shows $\xi_V(\tau)$ (relative to the undamped case $ b_0^2 
\exp(-\omega_0 \tau)/4 $ ) and Fig. \ref{fig:r(t)} the remainder function 
$r(t)$ for the chosen values of the damping parameter $\gamma$.

\noindent
Note that one has a simple
result for the {\it imaginary} part (which is odd in $t$)
\be
{\rm Im} \, \xi_V(it) \E - \frac{1}{4 m \Omega} \, \sin \left ( \Omega t \right ) \, 
e^{- \gamma |t|/2 }
\label{Im xiv}
\ee
but that the real part is more involved. This can also be seen if we use 
the standard representation of the step function
\be
\Theta(E) \E - \frac{1}{2 \pi i} \, \int_{-\infty}^{+\infty} ds \> 
\frac{\exp(-i s E)}{s + i \epsilon}
\ee
to extend the integration range in eq. (\ref{A:xiv 2}) to $ \> -\infty \> $. 
Using the identity
(\ref{antisym BW}) simple manipulations then give
\be
\xi_V(i t) \E \frac{1}{4 m \pi \Omega} \,  \int_{-\infty}^{+\infty} ds \>
\frac{\sin(\Omega s)}{s - t + i \epsilon} \, e^{- \gamma |s|/2}
\> \> , \> \> t \> \mbox{real} 
\ee
which shows that the real part is determined by a principal-value integral
whereas the imaginary part is given by eq. (\ref{Im xiv}).

Finally we consider the behaviour of the Vineyard function for small and large 
$\tau$. From eq. (\ref{A:xiv 4}) one immediately obtains
\be
\xi_V(\tau) \E \sum_{k=0}^{\infty} \, a_k \, \tau^k +  \sum_{k=1}^{\infty} \, 
b_{2k} \, \tau^{2k} \, \ln \left ( \omega_0 \tau \right )
\label{A:xiv low tau}
\ee
with
\bea
a_0 \! \EA \! \frac{1}{2 m \pi \Omega} \, \arctan \left ( \frac{2 \Omega}{\gamma} 
\right ) \> \> , \hspace{0.5cm} a_1 \E - \frac{1}{4 m} \> \> , \non
a_2 \! \EA \!  \frac{1}{4 m \pi \Omega}  \left [  \gamma \Omega \left ( 
\frac{3}{2} - \gamma_E \right ) + \left ( \Omega^2 - \frac{\gamma^2}{4} 
\right ) \arctan \left ( \frac{2 \Omega}{\gamma} 
\right )  \right ] , a_3 = - \frac{\gamma}{48 m} 
\left ( \Omega^2 - \frac{3\gamma^2}{2} \right ) 
\label{A:coeff}\\
\vdots && \non
b_2 \! \EA \! - \frac{\gamma}{4 m \pi}  \> \> , \hspace{0.5cm} b_4 \E
 - \frac{\gamma}{24 m \pi} \, \left ( \Omega^2 - \frac{\gamma^2}{4} \right ) 
\> \> , \> \> \ldots \> \> . \nonumber
\eea
Another possibility is to use the differential eqaution
\be
\xi_V^{IV}(\tau) -  \left ( 2 \omega_0^2 - \gamma^2 \right ) \, \xi_V''(\tau)
+ \omega_0^4 \, \xi_V(\tau) \E \frac{\gamma}{2 m \pi} \, \frac{1}{\tau^2}
\label{diff eq}
\ee
with the appropriate boundary conditions which follows directly from 
eq. (\ref{A:xiv 2}).

The asymptotic behaviour for arbitrary $\tau$ is more involved. Eq. 
(\ref{A:xiv 2}) 
may be used in the right-hand $\tau$-plane: one simply has to expand $\rho(E)$ 
for small $E$ and integrate term by term to find
\be
\xi_V(\tau) \> \stackrel{\tau \to \infty}{\longrightarrow} 
\> \frac{\gamma}{2 m \pi \omega_0^4} \, 
\frac{1}{\tau^2} + \frac{6 \gamma}{m \pi \omega_0^8} \, \left ( \Omega^2 - 
\frac{\gamma^2}{4} \right ) \, \frac{1}{\tau^4} + \ldots \hspace{0.5cm} 
{\rm Re} \> \tau \ge 0 \> .
\label{asy 1}
\ee
This is consistent with the result from
the differential equation (\ref{diff eq}) for large $\tau$ if one considers the
derivatives as corrections.
To find the asymptotic behaviour in the left-hand $\tau$-plane we may use the 
explicit representation (\ref{A:xiv 4}) and re-introduce the exponential integral
as this function has a simple asymptotic behaviour (see, e.g. eq. 5.1.51 in 
ref. \cite{Handbook})
\be
E_1 (z) \> \sim \> \frac{e^{-z}}{z} \, \left [ \, 1 - \frac{1!}{z} +  
\frac{2!}{z^2} - \ldots \, \right ] \> \> , \> \> |{\rm arg} \, z| < 3 \pi/2 \> .
\label{E1 asy}
\ee
However, care is needed when replacing Ein($z$) by the exponential integral since
the correct addition theorem of the logarithm with complex arguments  
\be
\ln \left ( a b \right ) \E \ln (a) + \ln (b) + 2 \pi i \, \Bigl [ \, 
\Theta(-{\rm Im} \, a) 
\Theta(-{\rm Im} \, b) \Theta({\rm Im} \, (ab)) - \Theta({\rm Im} \, a) 
\Theta({\rm Im} \, b) \Theta(-{\rm Im} \, (ab))\, \Bigr ] 
\ee
has to be used for $ \ln ( E_{r,s} \tau)$. Consequently
\be
\xi_V(\tau) \E - \frac{1}{8 m \pi \Omega} \sum_{r,s = \pm 1} \, r \, 
e^{E_{r,s} \tau} \, 
\Bigl [ \, 2 \pi \Theta (s {\rm Im} \, \tau) \, \Theta (- s {\rm Im} 
(E_{r,s}\tau)) 
+ i s \, E_1 \left ( E_{r,s} \tau \right ) \,\Bigr ] \> \> , \> \> 
| {\rm arg} \, \tau | < \pi \> .
\label{A:xiv 5}
\ee
Although eq. (\ref{A:xiv 5}) is less suited to display the analytic structure 
of the Vineyard function it allows to find the Stokes lines for the
asymptotic behaviour. These are the rays in the complex $\tau$-plane which divide 
the power-like decrease of eq.  (\ref{asy 1}) from an 
exponential increase. Indeed, due to eq.  (\ref{E1 asy}) the last term in the 
square brackets of eq. (\ref{A:xiv 5}) gives
rise to the  power-like decrease of $\xi_V(\tau)$ but this will 
be overwhelmed by the exponential increase of the first term if
\be
\Theta \left ( {\rm Re} (E_{r,s}\tau) \right ) \,\Theta (s {\rm Im} \, \tau) \, 
 \Theta \left ( - s {\rm Im} (E_{r,s}\tau) \right ) \> \ne \> 0 
\label{cond exp}
\ee
for any $r,s = \pm 1 $. Writing
\be
E_{1,1} \E \Omega + i \frac{\gamma}{2} \E \omega_0 \, e^{i \varphi_1} \> \> , \> 
\> \>  \varphi_1 \E \arctan \left ( \frac{\gamma}{2 \Omega} \right ) \> \> , 
\hspace{0.4cm} 0 \le \varphi_1 \le \frac{\pi}{2}
\ee
a straightforward analysis of the condition (\ref{cond exp}) shows that an 
exponential increase only occurs for
\be
\left | {\rm arg} \> \tau \right | \> > \> \frac{\pi}{2} + \varphi_1 \> ,
\ee
i.e. inside a sector around the cut with opening angle $ \pi/2 - \varphi_1 = 
\arctan(2 \Omega/\gamma)$. This means that eq. (\ref{asy 1}) holds for the 
wider range $ \, | {\rm arg} \> \tau | <  \pi/2 + \varphi_1 \, $.

\vspace{0.5cm}
\section{Asymptotic behaviour of the structure function}
\setcounter{equation}{0}
\label{app:B}

Here we derive the asymptotic behaviour of the structure function when the energy
transfer $\nu$ becomes very large. This is done by standard asymptotic analysis:
for example, one may apply eq. (30) in ref. \cite{Light} to our eq. 
(\ref{S inel 2}). Then one obtains
\be
S(q,\nu) \> \stackrel{\nu \to \infty}{\longrightarrow} \>  
\frac{2}{\pi} \, \left [ \> - \frac{1}{\nu} \, {\rm Im} \, \Phi_q(0)
+ \frac{1}{\nu^3} \, {\rm Im} \, \Phi_q''(0) - \frac{1}{\nu^5} \, 
{\rm Im} \, \Phi_q^{IV}(0) + \ldots \> \right ] \E \frac{\gamma}{m \pi} \,
\frac{q^2}{\nu^3} + \ldots
\label{S asy LO}
\ee
since $ {\rm Im} \, \Phi_q(0) = 0 $. Here we have used the representation of
the characteristic function in terms of the Vineyard function $\xi_V(it)$  and 
the low-$t$ expansion of the latter. Note that the leading 
contribution comes from the logarithmic term in
eq. (\ref{xiv small t}) which produces an imaginary part for $\xi_V''(0)$. 

However, higher-order terms cannot be calculated by means of eq. 
(\ref{S asy LO}) since ${\rm Im} \, \Phi_q^{IV}(0)$ does not exist. This 
shows that the next-to-leading asymptotic term is {\it not} falling off like
$1/\nu^5$. To determine this term we use the exponential representation
(\ref{S from Phi}), well-known results for the Fourier transform of generalized 
functions and the low-$t$ behaviour of the Vineyard function. Writing
\be
2 q^2 \, \left [ \, \xi_V(it) - \xi_V(0) \, \right ] \> =: \>  f(t) + g(t) \, 
\ln \left ( i \omega_0 t \right ) \E f(t) + g(t) \, \left [ \, \ln \left ( 
\omega_0 |t| \right ) + i \frac{\pi}{2} \,  {\rm sgn} \, t \> \right ]
\ee 
with
\be
f(t) \> \stackrel{t \to 0}{\longrightarrow} \> - i \frac{q^2}{2m} \, t + {
\cal O} \left ( t^2 \right ) \> \> ,
\hspace{1cm} g(t)  \> \stackrel{t \to 0}{\longrightarrow} \> \frac{\gamma}{\pi} 
\frac{q^2}{2m} \, t^2 +  {\cal O} \left ( t^4 \right )  
\ee
one simply gets by expanding the exponential
\bea
\Phi_q(t)  &  \stackrel{t \to 0}{\longrightarrow}  & 1 + f(t) + g(t) \, 
\left [ \, \ln \left ( \omega_0 |t| 
\right ) + i \frac{\pi}{2} \, {\rm sgn} \, t \> \right ] \non
&& + \frac{1}{2} f^2(t) 
+ f(t) \, g(t) \, \left [ \, \ln \left ( \omega_0 |t| 
\right ) + i \frac{\pi}{2} \, {\rm sgn} \, t \> \right ] 
\label{Phi small t} \\ 
&& + \frac{1}{2} g^2(t) \, \left [ \, \ln \left ( \omega_0 |t| 
\right ) + i \frac{\pi}{2} \, {\rm sgn} \, t \> \right ]^2 + \ldots \> \> . 
\nonumber 
\eea
As the Fourier transform of powers of $t$ gives derivatives of 
$\delta$-functions
we see that all regular terms do not contribute to the 
asymptotic behaviour for large $\nu$. The contribution of the 
non-analytic terms can be taken from Table 1 of ref. \cite{Light}
(setting $ y = - \nu/(2 \pi) $ )
\bea
\int_{-\infty}^{+\infty} dt \> t^n \, {\rm sgn} \, t \, e^{i \nu t} \EA 
2 \frac{n!}{(-i \nu)^{n+1}} \\
\int_{-\infty}^{+\infty} dt \> t^n \, \ln |t| \, e^{i \nu t} \EA i \pi
\frac{n!}{(-i \nu)^{n+1}} \, {\rm sgn} \, \nu  \> .
\eea
One then realizes that the leading contribution in the asymptotic expansion 
arises from the last term in the first 
line of eq. (\ref{Phi small t})
\be
S(q,\nu)  \> \stackrel{\nu \to \infty}{\longrightarrow} \> \frac{1}{2 \pi} \, 
\frac{\gamma q^2}{2 m \pi} \, 
\int_{-\infty}^{+\infty} dt \> t^2 \, \left [ \, \ln \left ( \omega_0 |t| 
\right ) + i \frac{\pi}{2} \, {\rm sgn} \, t \> \right ] \, e^{i \nu t}
\E \frac{\gamma}{m \pi} \, \frac{q^2}{\nu^3} \> ,
\ee
in agreement with eq. (\ref{S asy LO}). The subleading contribution stems from the 
last term in the second line
\be
\Delta S(q,\nu) \> \stackrel{\nu \to \infty}{\longrightarrow} \> 
\frac{1}{2 \pi} \, \left ( -i \frac{q^2}{2 m} \right ) 
\frac{\gamma q^2}{2 m \pi} \, 
\int_{-\infty}^{+\infty} dt \> t^3 \, \left [ \, \ln \left ( \omega_0 |t| 
\right ) + i \frac{\pi}{2} \, {\rm sgn} \, t \> \right ]  \, e^{i \nu t}
\E \frac{3 \gamma}{2 m^2 \pi} 
\, \frac{q^4}{\nu^4} \> ,
\label{S asy NOL}
\ee
whereas one can show that the last line gives a contribution of order 
$ (\ln \nu)/\nu^5 $. The
asymptotic expansion makes sense if the subleading term is much smaller than the 
leading one which requires $ \nu \gg q^2/(2m) $, i.e. excitation energies much 
larger than the maximum of the quasielastic peak.


\begin{thebibliography}{99}

\bibitem{HO} O. W. Greenberg, Phys. Rev. D {\bf 47}, 331 (1993);
S. A. Gurvitz and A. S. Rinat, Phys. Rev. C {\bf 47}, 2901 (1993);
E. Pace, G. Salm\`e and A. S. Rinat , Nucl. Phys. A {\bf 572}, 1 (1994).

\bibitem{smearing}
E. Pace, G. Salm\`e and F. M. Lev, Phys. Rev. C {\bf 57}, 
2655 (1998);
N. Isgur,  S. Jeschonnek, W. Melnitchouk and J. W. Van Orden, 
Phys. Rev. D {\bf 64}, 054005 (2001);
M.~W.~Paris and V.~R.~Pandharipande,
Phys.\ Rev.\ C {\bf 65}, 035203 (2002);
S. Jeschonnek and J. W. Van Orden, Phys. Rev. D {\bf 65}, 
094038 (2002);
M. W. Paris, nucl-th/0305020.

\bibitem{WeWi} H. A. Weidenm\"uller and A. Winther, Ann. Phys. (N.Y.) {\bf 66}, 
218 (1971).

\bibitem{HLM} Y. Horikawa, F. Lenz and N.C. Mukhopadhyay,
Phys. Rev. C {\bf 22}, 1680 (1980).

\bibitem{Weiss} U. Weiss: {\it Dissipative quantum systems}, 2nd edition, 
World Scientific (1999).

\bibitem{FeVe} R. P. Feynman and F. L. Vernon, Ann. Phys. (N.Y.) {\bf 24}, 118 
(1963).
'

\bibitem{CaLe} A. O. Caldeira and A. J. Leggett, Ann. Phys.  (N.Y.) {\bf 149}, 
374 (1983).

\bibitem{HMW} P. Chr. Hemmer, L. C. Maximon and H. Wergeland, Phys. Rev. {\bf 111}, 
689 (1958)

\bibitem{pol}  R. P. Feynman, Phys. Rev. {\bf 97}, 660 (1955); 
K. Mitra, A. Chatterjee and S. Mukhopadhyay, Phys. Rep. {\bf 153}, 91 (1987).

\bibitem{foot1} The definition of the 
characteristic function $\Phi_q(t)$ follows Appendix A of ref. \cite{WC6} which 
in the present context is more convenient than the original
$F(t)$ from ref. \cite{quasi}. One has $ \Phi_q(t) = F(-t)$ and thus,
in particular, Im $\Phi_q(t) = - {\rm Im} \> F(t)$.

\bibitem{WC6}
N. Fettes and R. Rosenfelder, Few-Body Syst. {\bf 24}, 1 (1998).

\bibitem{quasi} R. Rosenfelder, Ann. Phys.  (N.Y.) {\bf 128}, 188 (1980).

\bibitem{ChLi} T.-P. Cheng and L.-F. Li: {\it Gauge theory of elementary 
particle physics}, Oxford University Press (1988), chapter 1.2. 

\bibitem{Ing} G.-L. Ingold, in: {\it Coherent evolution in noisy environments},
A. Buchleitner and K. Hornberger (eds.), Lecture Notes in Physics, vol. {\bf 611}, 
p. 1, Springer (2002) [quant-ph/0208026].

\bibitem{foot2} A standard parametrization is
$J_{\> \rm Drude} = m \gamma \omega \, \omega_D^2/(\omega^2 + \omega_D^2) $
with a Drude cut-off $\omega_D$. This leads to 
$\gamma_{\> \rm Drude} = \gamma \,  \omega_D/(E + \omega_D) $. 
Also ``super-ohmic'' forms have been considered \cite{AoHo}.

\bibitem{AoHo} K.-I. Aoki and A. Horikoshi, Phys. Rev. A {\bf 66}, 042105 (2002) 
[quant-ph/0205002].

\bibitem{Vine} G. H. Vineyard, Phys. Rev. {\bf 110}, 999 (1958).

\bibitem{foot3} Given the common background it is not surprising that 
eq. (\ref{T eucl}) exhibits a close similarity with results from the polaron 
variational method \cite{WC1+7}.
Indeed, the exponent is proportional to the euclidean ``pseudotime'' 
$\mu^2(\tau)$ with 
a specific ``profile function''  $ A(E) = 1 + \gamma/|E| + \omega_0^2/E^2 $ 
which is singular for $E \to 0$ . Note also that the result is valid for all
space dimensions $d$ and not only for $d = 1$.

\bibitem{WC1+7} 
R. Rosenfelder and A. W. Schreiber, 
Phys. Rev. D {\bf 53} (1996) 3337;
Eur. Phys. J. C {\bf 25}, 139 (2002).

\bibitem{damped ho} 
H. Feshbach and Y. Tikochinsky, Transact. N. Y. Acad. Sci. {\bf 38}, 44 (1977);
D. Chru\'sci\'nski, math-ph/0209008; 
G. Vitiello, hep-th/0110182.

\bibitem{friction} B. A. Arbuzov, Theor. Math. Phys. {\bf 106}, 249 (1996) .

\bibitem{timeas}
A. Bohm, M. Loewe, and B. Van de Ven, quant-ph/0212130;
D. Chru\'sci\'nski, math-ph/0301024.

\bibitem{WC3}  
A. W. Schreiber, R. Rosenfelder and C. Alexandrou, 
Int. J. Phys. E {\bf 5}, 681(1996).

\bibitem{HaZw} A. Hanke and W. Zwerger, Phys. Rev. E {\bf 52}, 6875 (1995).

\bibitem{giant} G. Baur and C. A. Bertulani, Phys. Rev C {\bf 34}, 1654 (1986);
 G. Baur and C. A. Bertulani, in: {\it The Response of Nuclei under Extreme
Conditions}, eds. R. A. Broglia and G. F. Bertsch, Plenum Publishing (1988), p. 343;
J. Z. Gu and H. A. Weidenm\"uller, Nucl. Phys. A {\bf 690}, 382 (2001).

\bibitem{foot4} This deficiency is, of course, well known; see e.g. the discussion
in ref. \cite{WeWi}, below eq. (4.5).

\bibitem{foot5} This is due to the fact that the microscopic Hamiltonian 
(\ref{CL model}) does not contain momentum-dependent interactions.

\bibitem{RiRo} A. S. Rinat and R. Rosenfelder, Phys. Lett. B {\bf 193}, 411 (1987).

\bibitem{resp} R. Rosenfelder, Nucl. Phys. A {\bf 377}, 518 (1982).

\bibitem{Handbook} M. Abramowitz and I. Stegun (eds.) : {\it Handbook of
mathematical functions}, Dover (1965).

\bibitem{Luke} Y. L. Luke: {\it The special functions and their approximations}, 
vol. II, Academic Press (1969).

\bibitem{Light} M. J. Lighthill: {\it Introduction to Fourier analysis and 
generalised functions}, Cambridge University Press (1964), chapter 4.


\end{thebibliography}
\end{document}